\documentclass{article}

\usepackage{arxiv}

\usepackage[utf8]{inputenc} % allow utf-8 input
\usepackage[T1]{fontenc}    % use 8-bit T1 fonts
\usepackage{hyperref}       % hyperlinks
\usepackage{url}            % simple URL typesetting
\usepackage{booktabs}       % professional-quality tables
\usepackage{amsfonts}       % blackboard math symbols
\usepackage{nicefrac}       % compact symbols for 1/2, etc.
\usepackage{microtype}      % microtypography
\usepackage{lipsum}		% Can be removed after putting your text content
\usepackage{tgtermes} % times font
\usepackage{imakeidx}
\usepackage[toc,page]{appendix}
\usepackage{setspace}
\usepackage{float}
\usepackage[caption = false]{subfig}
\usepackage[final]{graphicx}
 \makeindex[columns=3, title=Index,intoc]

\usepackage{ dsfont }
\usepackage{amsthm,amssymb,amsmath,layout}
\usepackage{enumerate}
\usepackage{comment}
\usepackage{cite}
\usepackage{tikz}
\usepackage{pgfplots}
\usepackage{capt-of}
\usepackage{geometry}

\usepackage{graphicx}

\usepackage{hyperref}

\usepackage{tikz}
\usetikzlibrary{decorations.pathreplacing,positioning, arrows.meta}
\usepackage{algorithm}
\usepackage[noend]{algpseudocode}

\usepackage{tikz}
\usetikzlibrary{positioning,arrows,calc}
\tikzset{
modal/.style={>=stealth',shorten >=1pt,shorten <=1pt,auto,node distance=1.5cm,
semithick},
unworld/.style={circle,draw,minimum size=0.5cm,fill=red!15},
world/.style={circle,draw,minimum size=0.5cm,fill=gray!15},
point/.style={circle,draw,inner sep=0.5mm,fill=black},
reflexive above/.style={->,loop,looseness=7,in=120,out=60},
reflexive below/.style={->,loop,looseness=7,in=240,out=300},
reflexive left/.style={->,loop,looseness=7,in=150,out=210},
reflexive right/.style={->,loop,looseness=7,in=30,out=330}
}

\usepackage{listings}
\usepackage{color}

\definecolor{codegreen}{rgb}{0,0.6,0}
\definecolor{codegray}{rgb}{0.5,0.5,0.5}
\definecolor{codepurple}{rgb}{0.58,0,0.82}
\definecolor{backcolour}{rgb}{0.95,0.95,0.92}

\lstdefinestyle{mystyle}{
	backgroundcolor=\color{backcolour},   
	commentstyle=\color{codegreen},
	keywordstyle=\color{magenta},
	numberstyle=\tiny\color{codegray},
	stringstyle=\color{codepurple},
	basicstyle=\footnotesize,
	breakatwhitespace=false,         
	breaklines=true,                 
	captionpos=b,                    
	keepspaces=true,                 
	numbers=left,                    
	numbersep=5pt,                  
	showspaces=false,                
	showstringspaces=false,
	showtabs=false,                  
	tabsize=2
}

\title{Understanding Epidemic Data and Statistics:\\
A case study of COVID-19}

\author{
  Amirhoshang Hoseinpour Dehkordi\\
  School of Computer Science\\
  Institute for Research in Fundamental Sciences, Tehran, Iran\\  
  \texttt{amir.hoseinpour@ipm.ir} \\
   \And
 Majid Alizadeh
  \thanks{Corresponding author} \\
  School of Mathematics, Statistics and Computer Science\\
  College of Science, University of Tehran, Tehran, Iran\\  
  \texttt{majidalizadeh@ut.ac.ir} \\
   \And
 Pegah Derakhshan\\
  School of Medicine\\
  Iran University of Medical Sciences, Tehran, Iran\\  
  \texttt{pegah.derakhshan1997@gmail.com} \\
   \And
 Peyman Babazadeh\\
  School of Engineering\\
   Islamic Azad University Central Tehran Branch, Tehran, Iran\\  
  \texttt{pynbbz@gmail.com} \\
   \And
 Arash Jahandideh\\
  Adhesive and resin department\\
  Iran polymer and petrochemical institute IPPI, Tehran, Iran\\
   Tehran, Iran\\  
  \texttt{a.jahandideh@ippi.ac.ir} \\
}

\begin{document}
\maketitle

\begin{abstract}
The 2019-Novel-Coronavirus (COVID-19) has affected 116 (By March 12) countries and out of more than 118,000 confirmed cases. Understanding the transmission dynamics of the infection in each country which affected on a daily basis and evaluating the effectiveness of control policies is critical for our further actions. To date, the statistics of COVID-19 reported cases show more than 80 percent of infected had a mild case of disease, while around 14 percent of infected experienced a severe one and about 5 percent are categorized as critical disease victims. Today's report (2020-03-12; daily updates in the prepared website) shows the confirmed cases of COVID-19 in China, South Korea, Italy, and Iran are 80932, 7869, 12462 and 10075; respectively. Calculating the total Case Fatality Rate (CFR) of Italy (2020-03-04), about 7.9\% of confirmed cases passed away. Compared to South Korea's rate of 0.76\% (10 times lower than Italy) and China's 3.8\% (50\% lower than Italy), the CFR of Italy is too high. There are some effective policies that yield significant changes in the trend of cases. The lockdown policy in China and Italy (the effect observed after 11 days), Shutdown of all non-essential companies in Hubei (the effect observed after 5 days), combined policy in South Korea and reducing working hours in Iran.
\end{abstract}

%\flushbottom
%\maketitle
% * <john.hammersley@gmail.com> 2015-02-09T12:07:31.197Z:
%
%  Click the title above to edit the author information and abstract
%
%\thispagestyle{empty}

%\noindent Please note: Abbreviations should be introduced at the first mention in the main text – no abbreviations lists. Suggested structure of main text (not enforced) is provided below.

\section*{Introduction}
\label{sec:Introduction}
Human coronaviruses (HCoV) which causes gastrointestinal and respiratory tract infections, were first introduced by the discovery of HCoV-229E and HCoV-OC43, from the nasal cavities of human patients with the common cold, in 1960s (Myint, 1994 \cite{myint1994human}; Tyrrell \& Bynoe, 1966\cite{tyrrell1966cultivation}). Other discovered human coronaviruses, which have involved serious respiratory tract infections, include SARS-CoV (2003), HCoV NL63 (2004), HKU1 (2005), MERS-CoV (2012), and the latest one, SARSCoV-2 (2019) resulting in Coronavirus disease (COVID-19)(Lim, Ng, Tam, \& Liu, 2016\cite{lim2016human}; Syed, 2020\cite{syed2020coronavirus}). The name refers to the morphology of the virus, when viewed under 2D transmission electron microscopy (large pleomorphic spherical particles with bulbous surface) and stems from the Latin word "corona", meaning "crown"(Goldsmith et al., 2004 \cite{goldsmith2004ultrastructural}). Concerning the risk factor, HCoVs vary significantly; from the relatively harmless ones (i.e., common cold) to the most lethal ones (MERS-CoV, with more than 30\% mortality rate in the infected) (Fehr \& Perlman, 2015\cite{fehr2015coronaviruses}). CoVs spread during cold seasons and cause colds with major symptoms, i.e., fever, sore throat, and less commonly pneumonia and bronchitis for the more aggressive strains. To date, there are no vaccines or antiviral drugs capable of preventing or treating HCoV infections (Fehr \& Perlman, 2015 \cite{fehr2015coronaviruses}; Forgie \& Marrie, 2009 \cite{forgie2009healthcare}; Liu et al., 2017 \cite{liu2017prevalence}).\\
To date, several outbreaks of coronavirus-related diseases have been reported. Severe acute respiratory syndrome- or SARS was the first coronavirus-related outbreak, starting in Guangdong, China, in November 2002, and spread to a total of 29 territories, including Hong Kong, Taiwan, Canada, Singapore, Vietnam, and the United States, within 9 months. It infected a total of 8,098 people and killed 774 worldwide  (Smith, 2006\cite{smith2006responding}). The second coronavirus-related outbreak happened in the Middle East in April 2012, officially named Middle East Respiratory Syndrome or MERS. This virus was first identified in a patient from Saudi Arabia, and later, MERS affected several other countries, including Saudi Arabia, South Korea, the United Arab Emirates, Jordan, Qatar, and Oman. Overall resulting with infections in 24 countries, with over 1,000 cases and over 400 deaths (Organization\cite{worldcase}). The outbreak of MERS happened again in South Korea, supposedly from a traveler from Middle East. It happened during May and July 2015, and infected a total of 186 individuals, with a death toll of 36 (Organization, 2015\cite{world2015middle}). After 3 years in August 2018, the next MERS outbreak happened in countries on the Arabian Peninsula, and resulted in almost 147 infected people and the death of 47. The MERS outbreak had been reported in Saudi Arabia, United Kingdom, and South Korea.  \\ 
In Dec 2019, a pneumonia outbreak was reported in Wuhan, China, and on Dec 31, it was attributed to a new strain of HCoV, first named as 2019-nCoV by World Health Organization (WHO), and later renamed to SARS-CoV-2 by the International Committee on Taxonomy of Viruses. Almost 2 weeks later, on Jan. 11, 2020, Chinese state media reported the first fatality from the new discovered virus, that led to infection of dozens more. As of Jan 20, multiple countries reported their first cases, including Japan, South Korea and Thailand. The first confirmed case in the United States came the very next day in Washington State. Continuing its spread, Coronavirus presence was confirmed throughout the month of February in the Philippines (Feb 2), France (Feb 14), Iran (Feb 21), and as reports started in Italy on Feb 23; many more Europeans countries followed suit, reporting their first confirmed cases. As of this writing, the coronavirus has affected 116 (By March 12) countries, out of more than 118,000 confirmed cases, around 4,000 people have lost their lives. With China, Italy, Iran, and South Korea experiencing the worst cases of outbreaks and showing no sign of alleviation, the 2019-2020 outbreak of COVID-19 is now officially recognized as a pandemic by WHO. An outbreak or epidemic often refers to a sudden increase in the occurrence of an infectious disease, in a particular time and place. Pandemics are near-global epidemic outbreaks, where multiple countries across the world are involved (Green et al., 2002 \cite{green2002epidemic}).\\
The mentioned rapid trend of spread prompts a lot of concerns and questions such as; "How fast is the virus spreading?", "which policies or efforts could control the disease better?", and  "what is the main difference of COVID-19 outbreak with pervious epidemics?" Fortunately, the daily case detection changes are available and can be tracked almost in real time on the website provided by authors (\url{http://iuwa.ir/corona/}). The aim of this study is to provide the transmission trend from China to other countries and to report the daily confirmed cases, case fatalities and surveillance in every countries from the first day of outbreak until March 4th. Also, to evaluate the effect of each government policies in controlling the outbreak of COVID-19.

\section*{Methods}
\label{sec:BasicStat}
\subsection*{Basic Statistics} 
\begin{figure}[ht]
\includegraphics[width=1.0\textwidth]{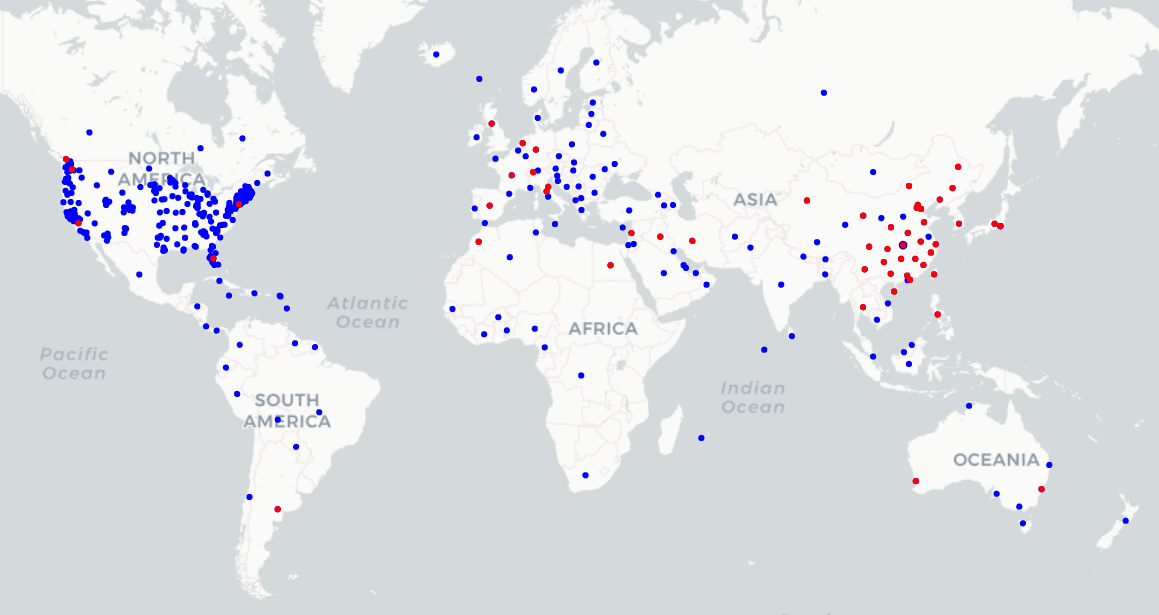}
\caption{Transmission of Coronavirus Disease 2019-2020 (COVID-19); Blue nodes represents regions with confirmed COVID-19 cases, and Red nodes represents the regions with COVID 19 causes deaths}
\label{fig:worldmap}
\end{figure}

COVID-19 has spread to 85 countries (By March 4) and most national authorities have failed to keep it's rapid spread contained\cite{githubJH}. WHO (World Health Organization) reports that it began in Wuhan city, located in Hubei Province of China, first reported on 21 January \cite{WHO1}. COVID-19 categorizes in three distinctions concerning it's infected host's severity of disease \cite{gong2020correlation} \cite{liu2020analysis}. To date, the statistics of its reported cases shows more than 80 percent of infected had a mild case of disease, while around 14 percent of infected experienced a severe one, suffering from breathlessness and pneumonia. 
And about 5 percent are categorized as critical disease patients, their symptoms include septic shock, respiratory failure, and the failure of more than one organ.\\
Reports on March 10 2020 show that China has the most confirmed, fatal and also recovered cases. The order of confirmed cases after China's, is followed respectively by South Korea, Italy, and Iran; which could be found in the table\ref{Tab:totCountry}.\\
\begin{table}[ht]
\caption{Top 10 total Confirmed,	Deaths \&	Recovered cases for March-6th}
\centering
  \begin{tabular}{l l l l}
    Country/Region &	Confirmed &	Deaths &	Recovered\\
    \hline
    China  &	80573 &	3042 &	 53888 \\
South Korea &	6593 & 	42 & 	135\\
Iran &	4747 & 	124 & 	913 \\
Italy &	4636 & 	197 & 	523 \\
Germany &	670 & 	0 & 	17\\
France &	653 & 	9 & 	12\\
Japan &	420 & 	6 & 	46\\
Spain &	400 & 	5 & 	2\\
US 	278 & 	14 & 	8\\
Switzerland &	214 &	1 & 	3\\
  \end{tabular}
  \label{Tab:totCountry}
\end{table}
Confirmed death cases caused by COVID-19 are also observed in 37 (By March 12) different countries, lead in numbers by China, Italy, Iran, and South Korea respectively. About 90\% of death cases are located in mainland China, where 80\% of confirmed cases were also present table\ref{Tab:totCountry}.  Vietnam, Cambodia, Nepal and Sri Lanka reported that all infected cases were recovered and there are no active cases that exist in their databases table\ref{Tab:totRecCountry}. 
\begin{table}[ht]
\caption{Countries with no active cases March-6th}
\centering
  \begin{tabular}{l l l l}
    Country/Region &	Confirmed &	Deaths &	Recovered\\
    \hline
    Vietnam  &	16 &	0 &	 16 \\
Macau &	10 & 	0 & 	10\\
Cambodia &	1 & 	0 & 	1 \\
Nepal &	1 & 	0 & 	1 \\
Sri Lanka &	1 & 	0 & 	1\\
  \end{tabular}
  \label{Tab:totRecCountry}
\end{table}
 The overall stats since March 6th states that there are 101104 confirmed, 3454 deaths and 55826 recovered cases, overall. Figure \ref{fig:worldmap} also shows that the COVID-19 spread exists in all continents.
\subsection*{Finding Linear Relations}
There isn't much known at the moment about COVID-19, so there is a small amount of data about its comprehensive effects and behaviors. In this study, relations are assumed to be linear, when, initially the drawn plot shows obvious linear relations and secondly, the fitted linear regression line had a small enough error to preserve the values given and the linear regression results could be interpreted with relative ease. In addition, by fitting regression lines with higher order, causes overfitting, resulting from the amount of data. There is no evidence yet, about the relation of other conditions to the outbreak and its case fatality rate (CFR), so by using linear regression line, policies and behaviors could be compared. In the prediction cases, by using linear regression, we could compare future trends of countries in earlier stages, with the ones in later stages. By considering above mentioned statements, we will find the best linear relation between arrays of data. In some cases, the linear relation could be observed but it may exhibit linear relation with some date shift of others (i.e. death cases should have linear relation with earlier values of confirmed cases, given the fact that it should take time from confirmation to death.)\\
CFR could be calculated by the following formula:
\begin{center}
$CFR=\dfrac{D_{eath}(T)}{C_{onfirmed}(T-dt)}$\\
\end{center}
In which $D_{eath}$ and $C_{onfirmed}$ functions will calculate the value of death cases and confirmed cases at that date, $T$ being the date we want to inspect the CFR, $dt$ is the mean duration of confirmed to death.

\begin{algorithm}
 To find linear relation with minimum error of two arrays $\mathcal{X},\mathcal{Y}$ of numbers, by shifting $\mathcal{Y}$ (maximum number of steps are $stp$) \;\\
\caption{BRG (Best Linear Regression) }\label{alg:BRG}
\begin{algorithmic}[1]
\Function{BRG}{$\mathcal{X},\mathcal{Y},stp, \eta$}\\
	\Comment{$\mathcal{N},x_0, \eta$ are NN, input point and neighborhood function respectively }
	\State $len \gets$ length$(\mathcal{X})-stp$ 
	\State $min_{val} \gets \inf$ 
	\State $best_{date} \gets 0$ 
	\Comment{length$(\mathcal{X})$=length$(\mathcal{Y})$}
	\State $rma \gets$ [] 
	\State $\mathcal{X'} \gets$ first $len$ elements of $\mathcal{X}$ 
	\For{$i \in \{1...stp\}$ }
		\State $\mathcal{Y'} \gets~ i-th$  $len$ elements of $\mathcal{Y}$
		\State $lm \gets  LinearRegression()$
        \State $   lm.fit(\mathcal{X'}, \mathcal{Y'})$ \Comment{Fitting best regression line to $\mathcal{X'},\mathcal{Y'}$ }
        \State mae $\gets$ lm.meanAbsoluteError
        \State rma.append(mae)
		\If{$mae < min_{val}$}
			\State $min_{val} \gets mae$
			\State $best_{date} \gets i$
			\State $best_{lm} \gets lm$
		\EndIf
	\EndFor
	\State \textbf{return} $best_{date}, best_{lm}$
\EndFunction
\end{algorithmic}
\end{algorithm}

\subsection*{Global Daily Statistics}
\label{sec:GlobalDailyStatistics}
\begin{figure}
\subfloat[ ]{\includegraphics[width =.5\linewidth]{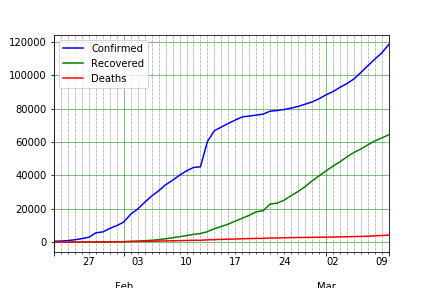}\label{fig:WorldConfirmed}} 
\subfloat[ ]{\includegraphics[width = .5\linewidth]{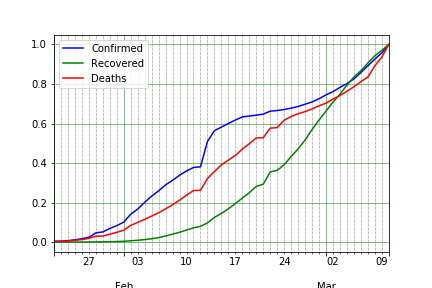}\label{fig:GlobaldiffConfirmedvsDeathsvsR}}\\
\subfloat[ ]{\includegraphics[width =.5\linewidth]{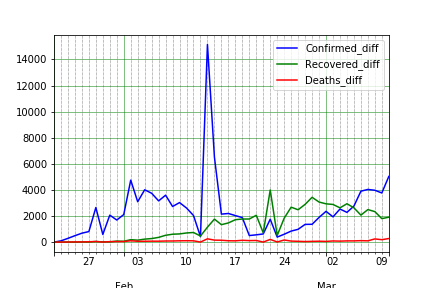}\label{fig:WorldNormalConfirmed}} 
\subfloat[ ]{\includegraphics[width = .5\linewidth]{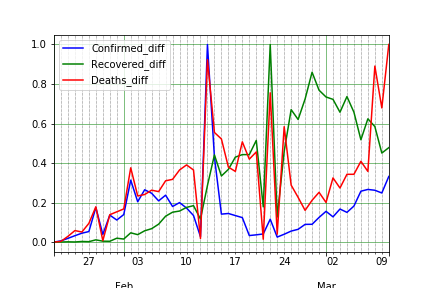}\label{fig:GlobalNormaldiffConfirmedvsDea}}\\
\caption{COVID-19 global epidemic data and statistics of. a) confirmed, recovered and death COVID-19 cases, b) normalized confirmed, recovered and death COVID-19 cases, c) New confirmed, recovered and death COVID-19 cases, and d) Normalized new confirmed, recovered and death COVID-19 cases}
\label{fig:GlobalData}
\end{figure}

Figure \ref{fig:WorldConfirmed} shows the global confirmed, deaths and recovered cases' trend for COVID-19 form Jan 22 to March 10, 2020. Death cases are excessively lower than confirmed ones, so we normalized (By dividing the value of Confirmed, Deaths and Recovered cases to their maximum respectively) it in Fig. \ref{fig:WorldNormalConfirmed} to investigate all three trends of cases. For the confirmed cases, there is a huge increase since Feb 11, the increase tones down from Feb 11 to the next day, Furthermore on Feb 13 another sharp increase is reported. It could be observed in Fig. \ref{fig:GlobaldiffConfirmedvsDeathsvsR} which shows new cases for each day (and normalized in Fig. \ref{fig:GlobalNormaldiffConfirmedvsDea} ). The most reliable speculation for this jump is that in that day, China (the country with the most confirmed cases) for the first time, reported the clinically diagnosed cases in addition to laboratory-confirmed cases \cite{WHO24}, in which 13332 clinically diagnosed cases are added to 1148 laboratory-confirmed ones. Since then, China has kept the same reporting method for the confirmed cases. On 23rd and 24th of Feb, again the decrease of new cases subsided and new cases increased as before. As shown in Fig. \ref{fig:WorldConfirmed} the reduction trend is continued (approximately) and the cause of the increase was other countries' growing numbers. So, for more accurate analysis each country will be investigated separately.

\section*{Results}
\subsection*{China Daily Statistics}
\label{sec:ChinaDailyStatistics}
Wuhan city located in Hubei Province is reported to be the origin of COVID-19. On January 23 2020, a lockdown in Wuhan and other cities in Hubei was implemented to control the outbreak of the COVID-19. A total of 12 other cities in Hubei, consisting of Huangshi, Jingzhou, Yichang, Xiaogan, Jingmen, Suizhou, Xianning, Qianjiang, Xiantao, Shiyan, Tianmen, and Enshi, restricted any form of transportation by the end of January 24. These decisions were made to prevent the expansion of COVID-19 any further.\subsubsection*{Confirmed Cases}
 As measured in \cite{backer2020incubation}, the incubation period for Wuhan travelers estimated from 2.1 to 11.1 days (The mean incubation period was estimated to be 6.4 days), and also generally the mean incubation period was estimated at 5.2 days which distributed in intervals of 4.1 to 7.0 with 95\% confidence \cite{li2020early}. By adding these two values, 11.6 days after Jan 24 (4th and 5th of Feb) the effects should be clearly manifesting.\\
Figure \ref{fig:HubeidiffConfirmedvsDeathsvsRe} depicted the new daily confirmed cases of China outside of Hubei. The peak of the plot is located on Feb 13 and the daily new cases reduces afterward. This reduction shows that the lockdown plays a serious role in the further reduction of cases in China(excluding Hubei province). Even Though there is no reason to argue the lockdown's positive impact on Hubei itself, the decrease in new confirmed cases (Feb 13 increase's rationale was described in the previous section) shows that emergency circumstances and movement limitations yields positive results to the reduction of confirmed cases from 10th of Feb. Since February 13 2020, the Chinese government has issued shutdown of all non-essential companies, including manufacturing plants, in Hubei Province. 5 days later, on Feb 18, a drop of new cases could be observed (Fig. \ref{fig:HubeidiffConfirmedvsDeathsvsRe}). Finally, the confirmed new cases in China were negligible from 1st of March. 
\subsubsection*{Deaths}
The number of deaths is far lower than confirmed cases. So, to investigate the relation of confirmed cases trend with the CFR, the normalized plot will be investigated. By observing normalized Hubei province plot of confirmed deaths and recovered cases in Fig. \ref{fig:NormalHubeiConfirmedvsDeathsvs}, it could be seen that the CFR trend behaves the same as the confirmation, with a shift (In date). Visually, it could be seen that the value of shift in date varies and increases during this time. In earlier cases, the period of confirmation cases leading to death was shorter. It seems, one reason for this variation is, that confirmed cases consist only of just laboratory cases, and by adding clinically diagnosed cases (which existed before but didn't count beforehand) the time of confirmation to death increases. In other words, the number of confirmed cases gets closer to the real value, and the cases are announced sooner than they did before. Other possible reasons include, the advancements in developing treatments, further delaying fatal cases, and the increase in public awareness, as more people with possible signs of infection come forward to be diagnosed.
 To estimate the expected value of confirming a case up to the death stage, assuming a linear relation between the death and confirmed rates, we draw a linear regression line for confirmed and death cases' value, Each time increasing the duration and finding the mean absolute error (MAE) of the regression line. Normally, by increasing the duration, following the reduction in investigated points, the MAE is reduced. However if there is an obvious relation between these two parameters, at the point which they had a correlating relation, MAE will begin to increase (Fig. \ref{fig:ChinaConfirmedDeathReg}). Wang et al. estimated the time from the the appearance of first symptoms to dyspnea was 5.0 days, to hospital admission 7.0 days, and to ARDS 8.0 days \cite{wang2020clinical}. Another study, found that the median days from the first symptom to death as 14 (range 6-41) days \cite{wang2020updated}. As seen in Fig. \ref{fig:ChinaNHConfirmedDeathReg} this value is about eleven in China (excluding Hubei province), Fig. \ref{fig:HubeiConfirmedDeathReg} depicted the value 5 for Hubei and Fig. \ref{fig:ChinaConfirmedDeathReg} depicted the value is 6 days for China. Assuming the hospital admission is the same day as confirmation (or a day before confirmation) the mean total of 14 days from the first symptom could be approved in\cite{wang2020updated}. By finding the mean day from confirmation to death, it is possible to find out the CFR in China, Hubei and China(excluding Hubei province) (algorithm \ref{alg:BRG} reports best shifting date value and best linear regression line). To find CFR of March 4th for Hubei, confirmed cases on March 4 should be divided by death cases of 5 days before that date (which is Feb 28) returning 4.4\%. Calculating CFR for China (excluding Hubei) on March 4, follows as confirmed cases on the same day divided by death cases of 11 days before that date (Feb 22) that equals 0.9\%; Finally, for China's CFR on March 4, dividing March 4th confirmed cases by death ones of 6 days prior (Feb 27) yields 3.8\%.\\

\subsubsection*{Recovered}
\begin{figure}
\subfloat[ ]{\includegraphics[width = .3\linewidth]{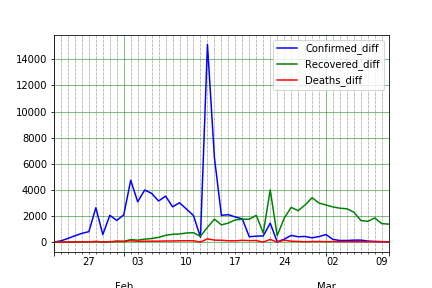}\label{fig:ChinadiffConfirmedvsDeathsvsRe}}
\subfloat[ ]{\includegraphics[width =.3\linewidth]{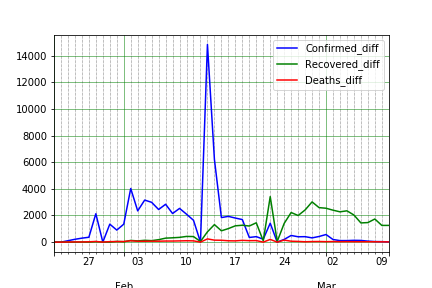}\label{fig:HubeidiffConfirmedvsDeathsvsRe}} 
\subfloat[ ]{\includegraphics[width = .3\linewidth]{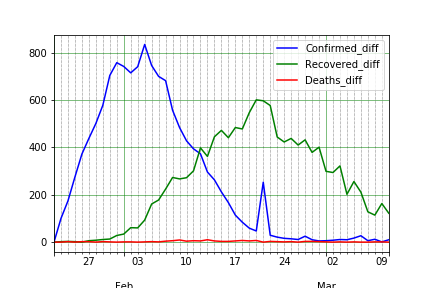}\label{fig:ChinaNHdiffConfirmedvsDeathsvs}}\\
\subfloat[ ]{\includegraphics[width =.3\linewidth]{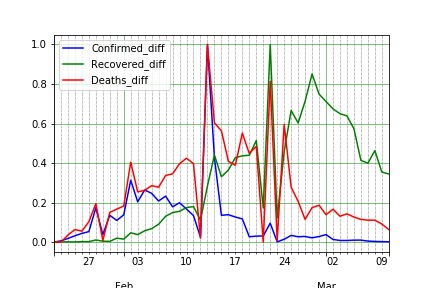}\label{fig:NormalChinadiffConfirmedvsDeat}} 
\subfloat[ ]{\includegraphics[width =.3\linewidth]{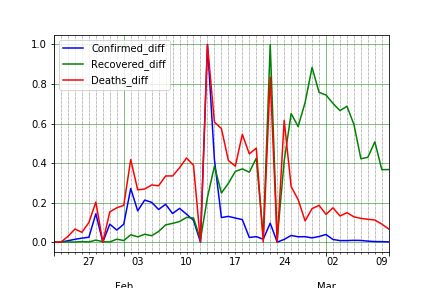}\label{fig:NormalHubeidiffConfirmedvsDeat}} 
\subfloat[ ]{\includegraphics[width =.3\linewidth]{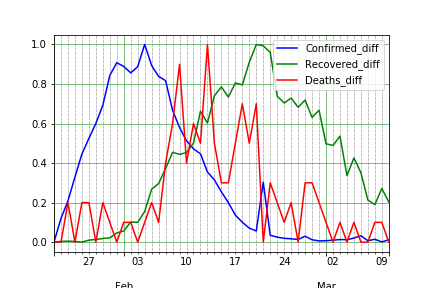}\label{fig:NormalChinaNHdiffConfirmedvsDe}} 
\\
\subfloat[ ]{\includegraphics[width =.3\linewidth]{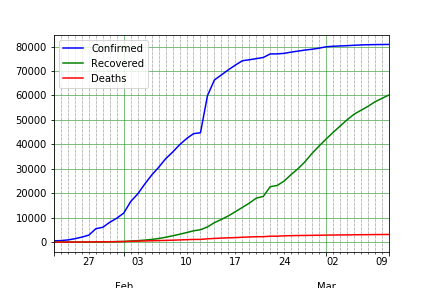}\label{fig:ChinaConfirmedvsDeathsvsRecove}} 
\subfloat[ ]{\includegraphics[width =.3\linewidth]{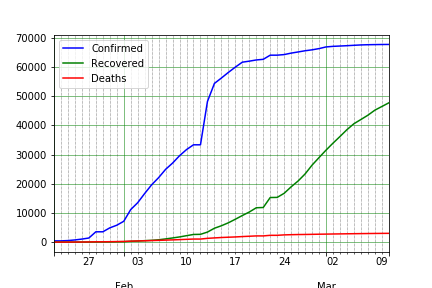}\label{fig:HubeiConfirmedvsDeathsvsRecove}}
\subfloat[ ]{\includegraphics[width =.3\linewidth]{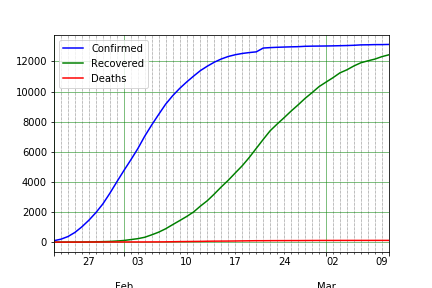}\label{fig:ChinaNHConfirmedvsDeathsvsReco}}\\ 
\subfloat[ ]{\includegraphics[width = .3\linewidth]{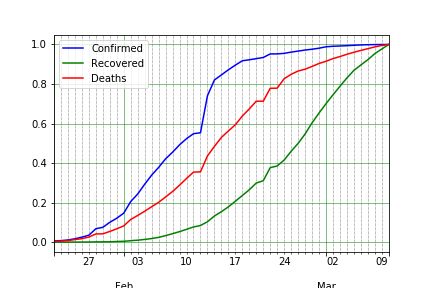}\label{fig:NormalChinaConfirmedvsDeathsvs}}
\subfloat[ ]{\includegraphics[width = .3\linewidth]{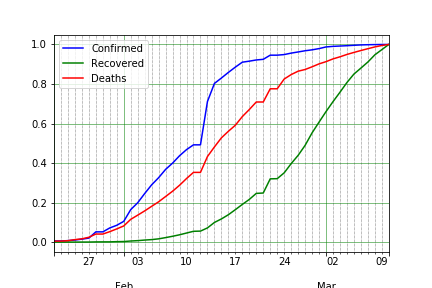}\label{fig:NormalHubeiConfirmedvsDeathsvs}}
\subfloat[ ]{\includegraphics[width = .3\linewidth]{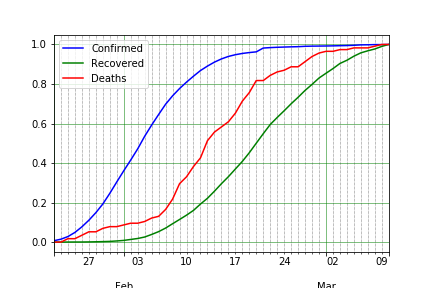}\label{fig:NormalChinaNHConfirmedvsDeaths}}\\
\caption{COVID-19 epidemic data and statistics of China (main land), Hubei province, and China excluding Hubei a-c) New [confirmed, recovered and death COVID-19 cases], d-f) Normal data [for confirmed, recovered and death COVID-19 cases], g-i) Confirmed and Death cases, and j-l) Normal data of confirmed and death cases}
\label{fig:ChinaData}
\end{figure}

\begin{figure}
\subfloat[ ]{\includegraphics[width =.5\linewidth]{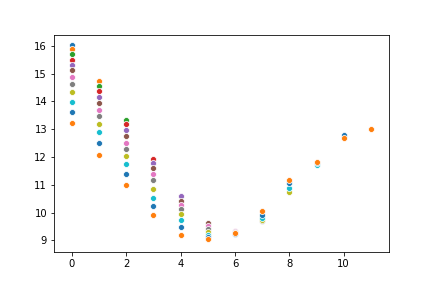}\label{fig:ChinaConfirmedDeathReg}} 
\subfloat[ ]{\includegraphics[width = .5\linewidth]{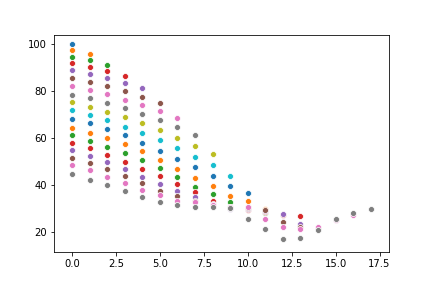}\label{fig:ChinaConfirmedRecoveredReg}}\\
\subfloat[ ]{\includegraphics[width =.5\linewidth]{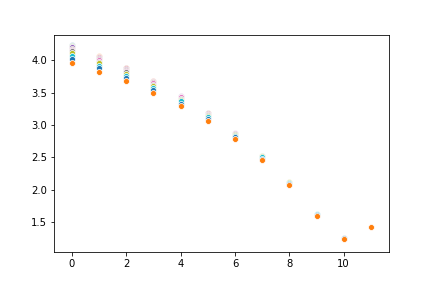}\label{fig:ChinaNHConfirmedDeathReg}} 
\subfloat[ ]{\includegraphics[width = .5\linewidth]{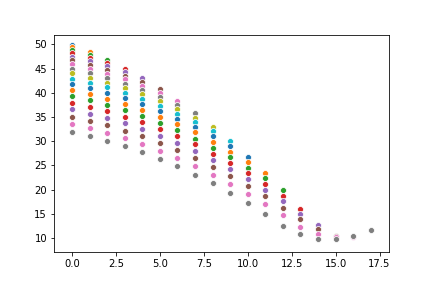}\label{fig:ChinaNHConfirmedRecoveredReg}}\\
\subfloat[ ]{\includegraphics[width =.5\linewidth]{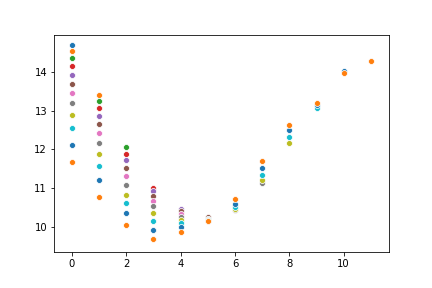}\label{fig:HubeiConfirmedDeathReg}} 
\subfloat[ ]{\includegraphics[width = .5\linewidth]{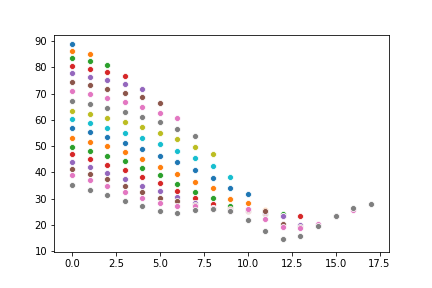}\label{fig:HubeiConfirmedRecoveredReg}}\\
\caption{Confirmed death and Confirmed recovered Regression MAE data for COVID-19 transmission through a,b) China (main land), c,d) Hubei province, and e,f) China excluding Hubei.}
\label{fig:ChinaData}
\end{figure}

Recovered cases are defined active cases-patients recovered after a certain amount of time, with its trend seen in Fig.  \ref{fig:NormalHubeiConfirmedvsDeathsvs}, Fig.
\ref{fig:NormalChinaConfirmedvsDeathsvs} and  Fig.
\ref{fig:NormalChinaNHConfirmedvsDeaths}
. By comparing recovered cases with confirmed ones, a relation is observed after date shifts. Unlike death cases, recovered cases' shift are longer initially, and reduces over time. The assumed reasoning is that by the passing of time, more medical treatments develop, healthcare providers gain more experience in handling patients' care and as more people are informed, increasing numbers of them check in hospitals at the early stages of their disease, resulting in an even more efficient treatment. However this reduction doesn't break the linear relation between confirmed cases and recovered ones enough to be significant. To find the mean date shift between confirmed cases and recovered ones, we apply a linear regression line to different dates by shifting them back until a first local minimum MAE is found(algorithm \ref{alg:BRG}). 
Hubei province's recovered mean duration value found is 12 days Fig.
\ref{fig:HubeiConfirmedRecoveredReg}, the same value for China(excluding Hubei province) is 14 days Fig. \ref{fig:ChinaNHConfirmedRecoveredReg}, and finally for China it's 12 days Fig.
\ref{fig:ChinaConfirmedRecoveredReg}. To find out the recovered cases ratio of Hubei province at March 4th, recovered cases of March 4 are divided by confirmed cases of 15 days prior resulting in 62.5\%. Same calculation applies for China (excluding Hubei) 92.9\%, and finally for China's result 67.4\%.

\subsection*{South Korea Daily Statistics}
\label{sec:SouthKoreaDailyStatistics}
\subsubsection*{Confirmed Cases}
\begin{figure}
\subfloat[ ]{\includegraphics[width =.25\linewidth]{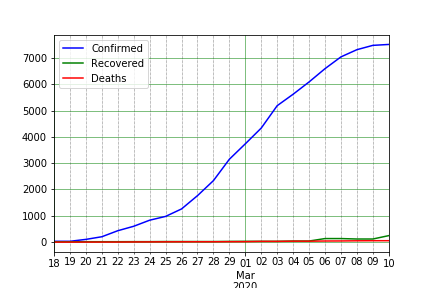}\label{fig:SouthKoreaConfirmedvsDeathsvsR}} 
\subfloat[ ]{\includegraphics[width = .25\linewidth]{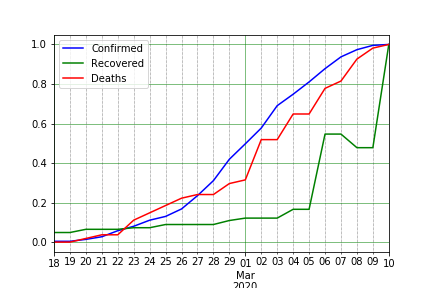}\label{fig:NormalSouthKoreaConfirmedvsDea}}
\subfloat[ ]{\includegraphics[width =.25\linewidth]{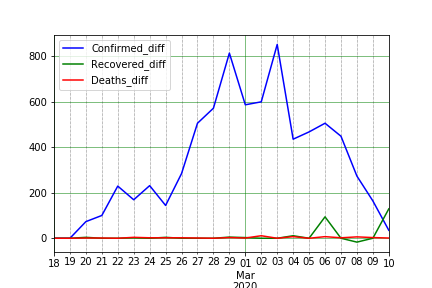}\label{fig:SouthKoreadiffConfirmedvsDeath}} 
\subfloat[ ]{\includegraphics[width = .25\linewidth]{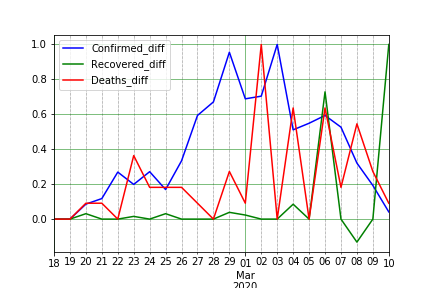}\label{fig:NormalSouthKoreadiffConfirmedv}}\\

\subfloat[ ]{\includegraphics[width =.25\linewidth]{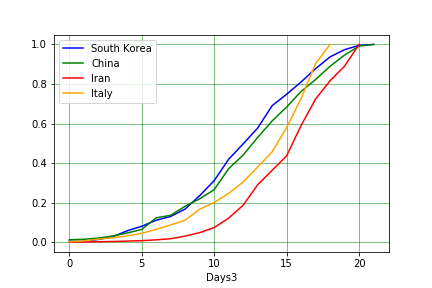}\label{fig:ConfirmedIran,Italy,SouthKorea}} 
\subfloat[ ]{\includegraphics[width = .25\linewidth]{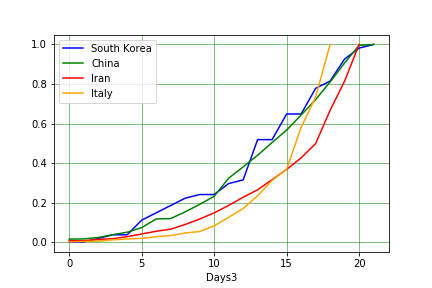}\label{fig:DeathsIran,Italy,SouthKorea,Ch}}
\subfloat[ ]{\includegraphics[width =.25\linewidth]{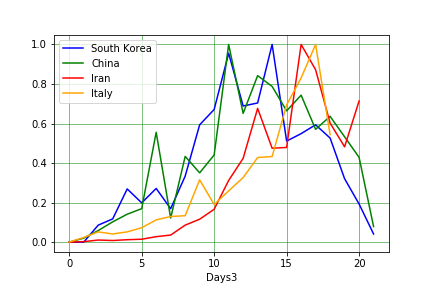}\label{fig:DiffConfirmedIran,Italy,SouthK}} 
\subfloat[ ]{\includegraphics[width = .25\linewidth]{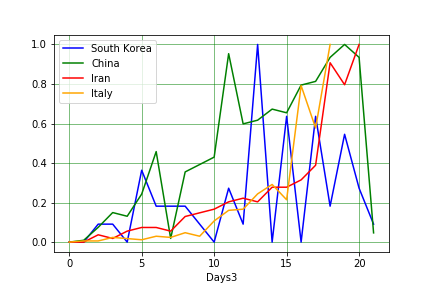}\label{fig:SouthKoreaDeaths_diffvsChinafi}}\\
\caption{COVID-19 epidemic data and statistics of South Korea [a-d], and between counties (South Korea, Iran, Italy, and China). For South Korea: a) confirmed, recovered and death COVID-19 cases, b) normalized data, c) New data, and d) Normalized new data. To compare between countries: e) Confirmed cases, f) Death cases, g) Confirmed rate, and i) New confirmed cases}
\label{fig:SouthKoreaData}
\end{figure}

First confirmed cases of COVID-19 was observed Jan 20th in South Korea, but the outbreak started around Feb 18th (29 days later), its death and recovered cases' trend shown in Fig.
\ref{fig:SouthKoreaConfirmedvsDeathsvsR}
 after normalizing, in addition the new cases' real-valued and normal form are also found in Fig. \ref{fig:NormalSouthKoreaConfirmedvsDea}
.
New confirmed, death and recovered cases are also depicted in Fig.
\ref{fig:SouthKoreadiffConfirmedvsDeath} and normalized in Fig. \ref{fig:NormalSouthKoreadiffConfirmedv}
.
In Fig.
\ref{fig:ConfirmedIran,Italy,SouthKorea}
first 16 day of four country's outbreak compared.
New confirmed and death cases also depicted in Fig. 
\ref{fig:DiffConfirmedIran,Italy,SouthK} and \ref{fig:SouthKoreaDeaths_diffvsChinafi}.\\
Being one of the first countries reporting the outbreak of COVID-19, the first report was on Jan 20. However, no outbreak was observed by Feb 18th, after which there was a significant increase in the number of confirmed cases. Comparison between the growth trend in numbers from South Korea from this date with China's in the early stages is depicted in Fig. \ref{fig:ConfirmedIran,Italy,SouthKorea} as shown the growth rate patterns are approximately the same between the two, showing a very similar behavior in the increasing numbers of confirmed cases. COVID-19 spreads from human to human\cite{chen2020covid}, therefore in addition to isolation of people, social avoidance and quarantine policies, faster detection of infected cases should reduce further growth. Each infected individual, by having contact with others directly or by proxy, (in other words by activity) could infect a number of people, so to reduce the odds of transmission of virus, faster detection of infected individuals could play a key role alongside lockdown strategies. \\
Based on the Korea Centers for Disease Control and Prevention (KCDC), as of March 8th 2020, a total number of 188,518 cases have been screened and tested for HCoV infection\cite{KCDC}. More than 10,000 tests had been taken on March 8th, and just in two days, this number has reached to 210,144 (indicating more than 20,000 tests had been taken in 2 days).  At this date, South Korea reached to 4099 taken-tests per million people. The magnitude of this number clearly shows the policy of South Korea, an attempt to reduce the duration of detection, through a faster detection strategy\cite{worldometersSKT}.
Furthermore, the fact that the CFR in the old South Korean affected adults\cite{KCDC1} (patients above 70 years) is still lower than that of Italian HCoV affected patients (average 47 years old), better signifies the role of early detection in controlling the epidemic.  
\subsubsection*{Deaths}
%CFRs shown in Fig.\ref{fig:DeathsIran,Italy,SouthKorea,Ch}, could be studied similarly to China's from above sections.\\
To determine the CFR of confirmed cases, specially in countries in which the COVID-19 is spreading, the knowledge of confirmation to death duration is essential. To demonstrate a wrong approach, if dividing the death cases by confirmed ones on a specific date, yield the wrong answers, for the death cases might have been confirmed some day prior. By assuming that CFR has a linear (or near linear) relation with confirmed rate, duration of confirmation to death could be evaluated by fitting multiple linear regression lines. Minimum value for MAE appears for 3 days shift, meaning death cases were confirmed 3 days before. To discern the CFR of South Korea, total death cases should be divided by total confirmed date of 3 days prior (slope of the regression line also shows the rate).\\
As alluded to before, there are many types of COVID-19 with respect to acuteness of the disease. The CFR varies depending on the level of infected people already confirmed. If a country manages to diagnose and confirm the affection of a patient, in an earlier stage and begins curating the infected individuals, or tallying those with mild COVID-19, the CFR would be comparatively lower. Such percussion alludes to the apparent lower CFRs in South Korea (77 out of 10,000) versus other countries with comparably large outbreaks.
\\
%It is not proven yet, what is the relation of race to define CFR, but if it is negligible the CFR could represent where is the threshold (or evaluation methods) of a country to confirm COVID-19 (if the CFR assumed to be fixed due to race what other parameters could play a role? race? health assets? GDP? any idea?).

\subsubsection*{Recovered}
Due to the small duration of outbreak in South Korea and larger duration of recovery date, it is difficult to accurately estimate the duration in this country. Considering the low CFR and a huge reduction of new cases in South Korea, the only remaining concern is the low rate of recovered cases.

\subsubsection*{Predictions}
Figure \ref{fig:ConfirmedIran,Italy,SouthKorea} displays the confirmed cases of South Korea with the ones in early stages of outbreak in China, with a comparably similar incremental trends (the relation was assumed to be linear relation between the two). to demonstrate this linear relation with a shift in date shift, we will follow the same  as before. By fitting a regression line, it could be estimated which date shift yield the lowest error (which is 0 date shift). Owing to the fact that China has experienced the COVID-19 after these fitting trends, it is possible to predict South Korea's next trend, assuming South Korea's policies are as good as China's (Since other conditions like race and climate are comparatively the same ). As mentioned before, China on Feb 13 (22 days after the outbreak), added clinical diagnosis cases to laboratory-confirmed cases, so it seems appropriate to be considered in predictions, after normalization.
Following the evaluation of the number of days shift, South Korea's confirmation rate behaves the same as China, resulting in the answer of 0.
 \\

\subsection*{Italy Daily Statistics}
\label{sec:ItalyDailyStatistics}
\subsubsection*{Confirmed Cases}
\begin{figure}
\subfloat[ ]{\includegraphics[width =.3\linewidth]{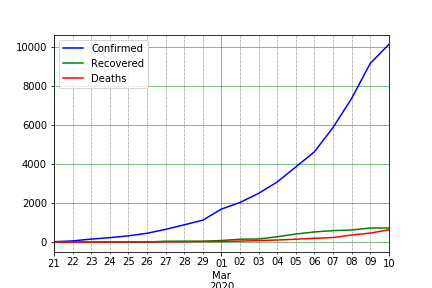}\label{fig:ItalyConfirmedvsDeathsvsRecove}} 
\subfloat[ ]{\includegraphics[width = .3\linewidth]{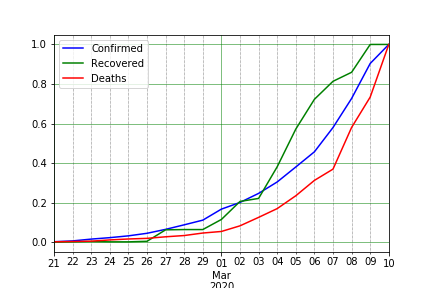}\label{fig:NormalItalyConfirmedvsDeathsvs}}
\subfloat[ ]{\includegraphics[width =.3\linewidth]{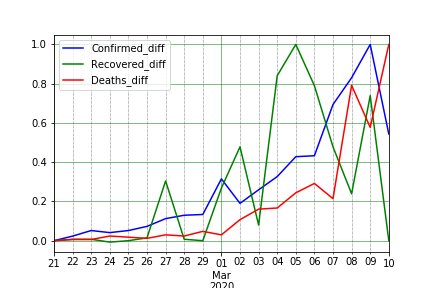}\label{fig:NormalItalydiffConfirmedvsDeat}} \\
\subfloat[ ]{\includegraphics[width =.3\linewidth]{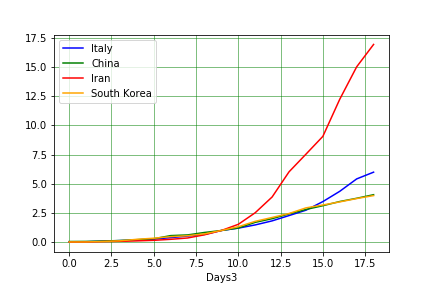}\label{fig:ConfirmedItalyvsChina}} 
\subfloat[ ]{\includegraphics[width = .3\linewidth]{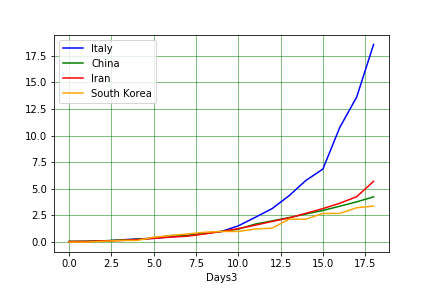}\label{fig:DeathsItalyvsChina}}
\subfloat[ ]{\includegraphics[width =.3\linewidth]{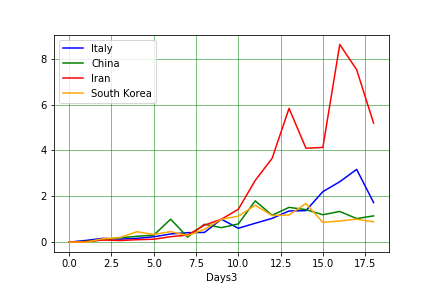}\label{fig:Confirmed_diffItalyvsChina}} \\
\subfloat[ ]{\includegraphics[width = .3\linewidth]{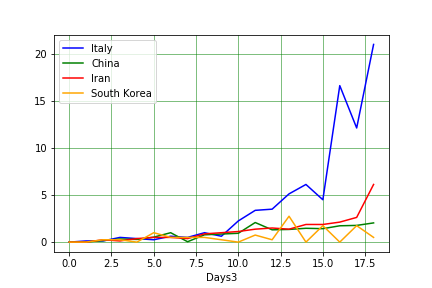}\label{fig:Deaths_diffItalyvsChina}}
\subfloat[ ]{\includegraphics[width =.3\linewidth]{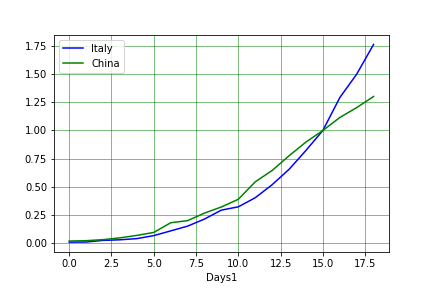}\label{fig:II-ConfirmedItalynotLombardyvs}} 
\subfloat[ ]{\includegraphics[width = .3\linewidth]{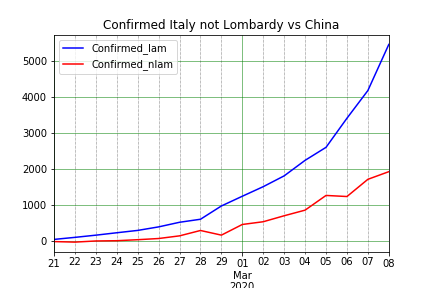}\label{fig:ConfirmedItalynotLombardyvsChi}}\\
\caption{COVID-19 epidemic data and statistics of Italy: a) confirmed, recovered and death COVID-19 cases, b) normalized data, c) Normalized new data. Comparison of COVID-19 statistics between China and Italy: d) Confirmed cases, e) Death cases, f) New confirmed cases, and g) New death cases. h) confirmed cases for Italy-excluding Lombardy and China, and i) death cases for Italy-excluding Lombardy and China.}
\label{fig:ItalyData}
\end{figure}

Italy reported its first confirmed cases of COVID-19 infection on Jan 31, later announcing an outbreak around Feb 21st (21 days later). We have displayed (after normalizing) overall analysis in addition to death and recovered cases' trend in Fig.
\ref{fig:NormalItalyConfirmedvsDeathsvs}
depicted the COVID-19's trend in the country. By normalizing the plot, death and recovered cases' trend could be seen in Fig.\ref{fig:NormalItalyConfirmedvsDeathsvs}.\\
The plot for new cases also figured in Fig.
\ref{fig:NormalItalydiffConfirmedvsDeat}
in real-valued and normal form. 
Comparison between the growth trend in numbers from Italy from this date with China's in the early stages is depicted in Fig. \ref{fig:ConfirmedItalyvsChina} as shown, the growth rate patterns are approximately the same between the two for first 14 days, showing identical behavior in the increasing numbers of confirmed cases. After day 14th, Italy exhibits incremental increase in confirmed cases juxtaposed to China's. March 10 (17th day of Italy's outbreak) being Italy's initial quarantine date, means it is too early to observe this policy's effects. Lombardy, the center of outbreak in Italy was locked down on February 22 and this lockdown strategy seems to have had a positive effect on other municipalities depicted in Fig. \ref{fig:ConfirmedItalynotLombardyvsChi} and by comparing other municipalities confirmed cases with China's (The mainland) trend shown in Fig.\ref{fig:II-ConfirmedItalynotLombardyvs} , results seem identical.\\
To analyze the stage of the disease in the Italian HCoV infected patients (i.e., to find out how rapid the infected people have been screened and diagnosed), the results of the COVID-19 tests of Italy and South Korea can be compared. South Korea has been considered as a reference as its HCoV-related CFR is low enough, indicating that both countries are at the same level of HCoV infection. Till March 10th 2020, a total of 60,761 HCoV tests had been taken form Italians- through their screening program, where as in South Korea this number exceeds 210,144 tests, more than 3 folds. However, the number of the daily taken tests in Italy has been increasing, i.e., it was 13000 on March 11th \cite{githubI}.
\subsubsection*{Deaths}
By comparing new cases of Italy and China which manifested in Fig. 
\ref{fig:DeathsItalyvsChina} (and the overall CFR comparison of the two in Fig.\ref{fig:Deaths_diffItalyvsChina}), it is clear that new death cases of Italy increased next to China's. Calculating the total CFR of Italy (confirmation to death duration calculated as 2 days) about 7.9\% of confirmed cases passed away.  Compared to South Korea's rate of 0.76\% (10\% lower than Italy) and China's 3.8\% (50\% lower than Italy), the CFR of Italy is too high. Ignoring race and climate as conditions (in which there is no clue of their impact), there should exist a strong rationale for this difference. One hypothesis is that some infected individuals are not diagnosed until later more serious stages of the disease. This could also explain the increase in confirmed cases, suggesting those infected, remain in contact with others. By comparing hospital beds per 10000 people, the indicator was 115.32 (At 2014), 42 (At 2012) and 34.22 (At 2012)(Iran 15 2014) for South Korea, China and Italy, respectively \cite{WHOHBPP}. 
 Statistically, it is observed that the CFR of COVID-19 has had a direct relation with age (with the average age of death for these 3 countries respectively 28, 38 and 47 years old \cite{worldometersP}) and It also potentially contributes to the increase in CFRs. 
\subsubsection*{Predictions}
Using the home quarantine strategy in Italy should fit the trend of China, with the decrease of new cases, at latest predicted to start on March 15. Comprehensively inspecting signs of the infection in people and faster starting treatment could also help lower CFR and make Italy's COVID-19 CFR more realistic.

\subsection*{Other Countries Daily Statistics}
\label{sec:OtherCountriesDailyStatistics}
\subsubsection*{Iran}
\begin{figure}
\subfloat[ ]{\includegraphics[width =.25\linewidth]{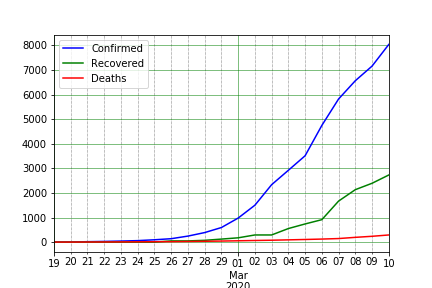}\label{fig:IranConfirmedvsDeathsvsRecover}} 
\subfloat[ ]{\includegraphics[width = .25\linewidth]{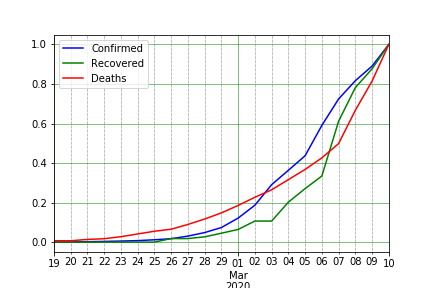}\label{fig:IranNormalConfirmedvsDeathsvsR}}
\subfloat[ ]{\includegraphics[width =.25\linewidth]{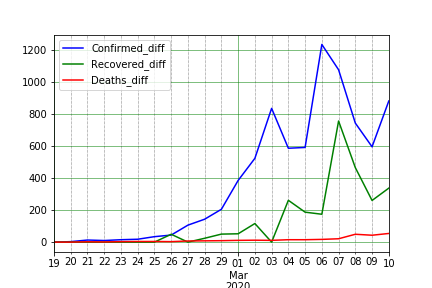}\label{fig:IrandiffConfirmedvsDeathsvsRec}} 
\subfloat[ ]{\includegraphics[width = .25\linewidth]{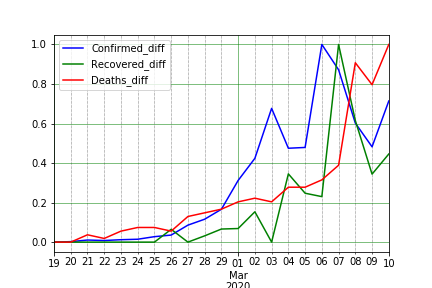}\label{fig:IranNormaldiffConfirmedvsDeath}}\\

\subfloat[ ]{\includegraphics[width =.25\linewidth]{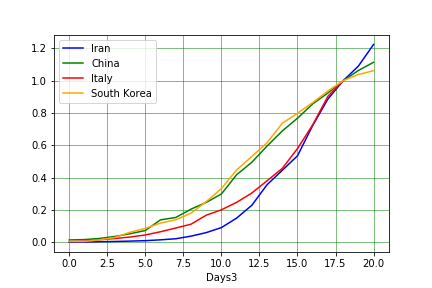}\label{fig:IranConfirmedvsChina}} 
\subfloat[ ]{\includegraphics[width = .25\linewidth]{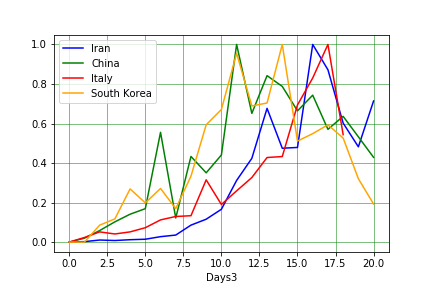}\label{fig:IranConfirmed_diffvsChina}}
\subfloat[ ]{\includegraphics[width =.25\linewidth]{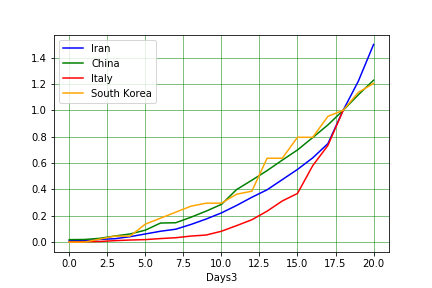}\label{fig:IranDeathsvsChina}} 
\subfloat[ ]{\includegraphics[width = .25\linewidth]{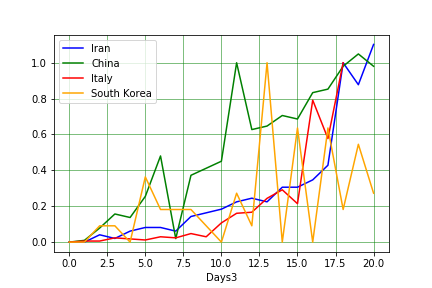}\label{fig:IranDeaths_diffvsChina}}\\

\subfloat[ ]{\includegraphics[width =.25\linewidth]{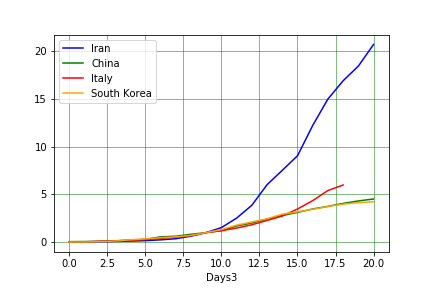}\label{fig:IranConfirmedvsChinafirst10}} 
\subfloat[ ]{\includegraphics[width = .25\linewidth]{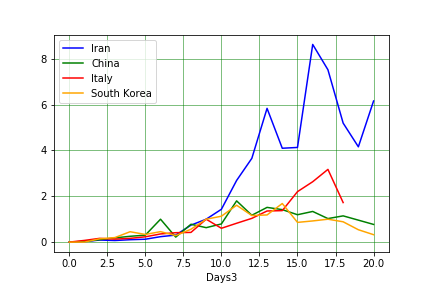}\label{fig:IranConfirmed_diffvsChinafirst}}
\subfloat[ ]{\includegraphics[width =.25\linewidth]{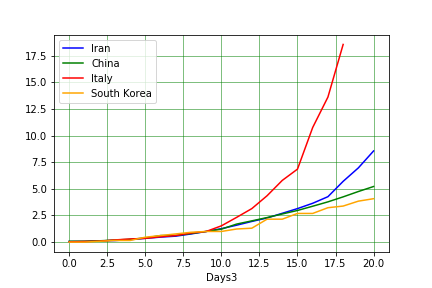}\label{fig:IranDeathsvsChinafirst10}} 
\subfloat[ ]{\includegraphics[width = .25\linewidth]{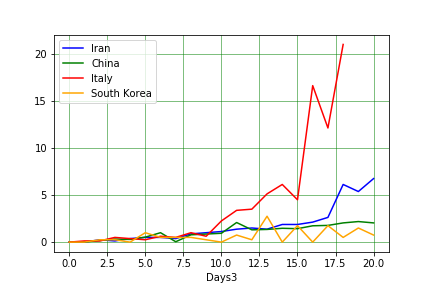}\label{fig:IranDeaths_diffvsChinafirst10}}\\
\caption{COVID-19 epidemic data and statistics of Iran: a) confirmed, recovered and death COVID-19 cases, b) normalized data, c) new data, and d) Normalized New data. Comparison of COVID-19 statistics between China and Iran: e) Confirmed cases, f) new data, g) death cases, and h) new death cases. Comparison between China and Iran in the first 10 days i) Confirmed cases, j) new confirmed cases, k) Death cases, and l) new death cases.}
\label{fig:IranData}
\end{figure}

COVID-19 was first reported in Iran starting with 2 dead cases on 19th of Feb in Qom (a city near Tehran, Iran's capital), followed immediately by a huge outbreak, displayed in Fig. \ref{fig:IranConfirmedvsDeathsvsRecover}  detailing its number of confirmed, deaths and recovered cases, and normalized in Fig. \ref{fig:IranNormalConfirmedvsDeathsvsR}.
Officially, Iran hasn't mandated any city wide lockdowns, but recently some provinces are refusing non-local travelers. Regarding the experiences in China, lockdown strategy will isolate cities to avoid transmitting the growth rate to other cities. However, such policies entail some negative consequences, and therefore they were not implemented in early stages of the outbreak in Iran; in contrast to Italy and South Korea, where there was a gap between outbreak and huge outbreak (21 and 27 days). On the 10th day of outbreak in Iran, total reported confirmed infected individual cases reached 388. \\
 To find confirmed to death cases mean duration, as before, we calculate the minimum MAE, assuming there is a linear relation between these two. Linear regression algorithms shows the date shift should be zero, which means on average, death cases confirmed at the same day. By comparing the death trend with confirmed cases in Fig. \ref{fig:IranNormalConfirmedvsDeathsvsR} and Fig. \ref{fig:IranNormaldiffConfirmedvsDeath}, it is obvious that the trend of CFR was not linear. The reported death vs confirmed are too excessive at early days.  It is possible that the shortage of COVID-19 testing kits or/and the lack of clinical diagnosis approach caused such unreliable data. Countries at this stage mostly had about 2 days of confirmation to death period, so following this assumption the COVID-19's CFR in Iran is around 4.4\%.  \\
By observing the 24th of Feb's report, there were 49 active cases (The cases which are not recovered or dead but confirmed as a COVID-19 infected) reported (at least 14 cases are new) and the value for active cases in the next day was 79 (at least 30 new cases). Two days later, the reported recovered cases was 49. Supposing no one was treated in less than two days, this results in 100\% recovery of 14 new cases in two days and also no active cases after two days, marking a great experience which was reported by Iran's officials. Otherwise, some cases were treated in one day. This makes Iran's phenomena intriguing for investigating the details and contributing factors for such cases in order to find a relation between individual's physical conditions and their treatment period. However, errors in report could also explain this information. Using regression line to find the mean recovered duration shows that the recovery duration for Iran should be around 1 day, Also seen in Fig. \ref{fig:IrandiffConfirmedvsDeathsvsRec}, the rate of recovery in Iran is quite significant at 38\% (keep in mind, Iran is in early stage of COVID-19's growth). \\
Figures \ref{fig:IranConfirmedvsChina}, \ref{fig:IranConfirmed_diffvsChina}, \ref{fig:IranDeathsvsChina} and \ref{fig:IranDeaths_diffvsChina} show Iran's trends of confirmed, newly confirmed, death and new death cases respectively. The trends display no linear relation, Therefore in order to find the linear relation, we fit Iran's with China's first 10 days of confirmed, newly confirmed, death and new death cases respectively depicted in Fig. \ref{fig:IranConfirmedvsChinafirst10}, \ref{fig:IranConfirmed_diffvsChinafirst}, \ref{fig:IranDeathsvsChinafirst10} and \ref{fig:IranDeaths_diffvsChinafirst10}.\\
The 10th day after the outbreak, was the day in which new cases of china started to decrease. in Contrast, Iran's incremental trend continued to rise until the 20th day of the outbreak, Even Though a temporal reduction in new cases could be observed starting on 6th of March persisting until 9th of March. Evidently, it is the impact of a nation wide implemented policy reducing working hours 6 days prior (5.4 days is the mean incubation period \cite{backer2020incubation} ). After 4 days, with the arrival of weekends and the following 2 additional holidays. Reports indicate, many people took trips during these dates which is widely assumed as a reason for confirmed cases' rise. Although many companies implemented forms of remote work for their employees (which is estimated to have a positive impact on preventing the growth outbreak), yet many governmental office working hours returned to their pre-outbreak times (naturally assumed to negatively impact the prevention of disease's spread ). There is yet no observable comparison between the effectiveness of either decisions, however the aggregation of effects could be seen from March 15. \\
An important upcoming event in Iran is Nowruz which is the Iranian New Year, when traditionally, gatherings are very common, which could have very serious consequences on the outbreak of COVID-19 in Iran, prompting immediate and decisive preventive policies from the government.

\subsubsection*{Japan}
\begin{figure}
\subfloat[ ]{\includegraphics[width =.5\linewidth]{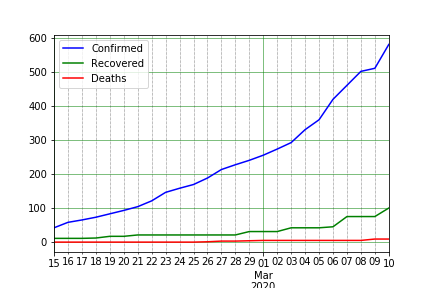}\label{fig:JapanConfirmedvsDeathsvsRecove}} 
\subfloat[ ]{\includegraphics[width = .5\linewidth]{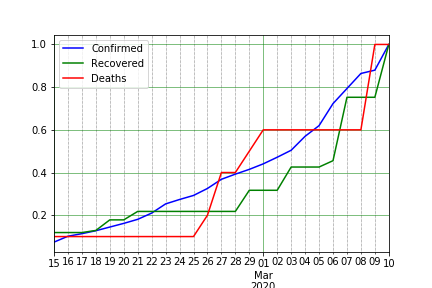}\label{fig:JapanNormalConfirmedvsDeathsvs}}\\
\subfloat[ ]{\includegraphics[width =.5\linewidth]{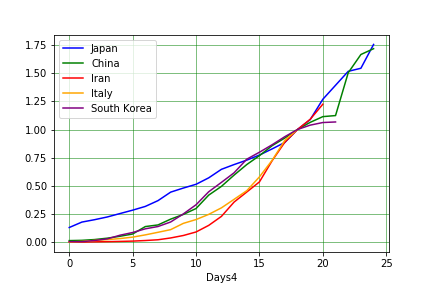}\label{fig:JapanConfirmedvsChinafirst17Ir}} 
\subfloat[ ]{\includegraphics[width = .5\linewidth]{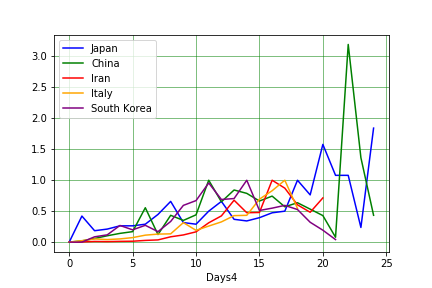}\label{fig:JapanConfirmed_diffvsChinafirs}}\\
\caption{COVID-19 epidemic data and statistics of Japan: a) confirmed, recovered and death COVID-19 cases, b) normalized data for confirmed, recovered and death COVID-19 cases, c) Confirmed cases of different countries and d) new confirmed cases of different countries.}
\label{fig:JapanData}
\end{figure}

The first confirmed cases of COVID-19 in Japan returned from Wuhan on 6th of Jan. 
%Since Feb 3, Japan has banned people with travel history to Wuhan and its passport holders,
 However, the outbreak in Japan sped up on Feb 15. Fig. \ref{fig:JapanConfirmedvsDeathsvsRecove} shows the trend of outbreak after 15th Feb in Japan, while Fig. \ref{fig:JapanNormalConfirmedvsDeathsvs} is normalized to compare Confirmed, Deaths and Recovered cases. The trend shows that the spread of COVID-19 in Japan behaves at a lower rate than China's first dates (see Fig. \ref{fig:JapanConfirmedvsChinafirst17Ir}). The number of deaths in Japan is also lower than (10 individuals on March 10) the regression line could be fitted with a high degree of confidence, However it is estimated that it is about 2\%. The CFR is very good by adding the information of mean age in Japan, which is 48. By finding out the hospital per bed value of Japan which is 134 (reported at 2012 WHO), the relation of death cases compared with hospital per bed is more clear.\\
 Although Japan managed the early stage exponential rate of COVID-19 (see Fig. \ref{fig:JapanConfirmedvsChinafirst17Ir}), a reduction in new cases is not observed (see Fig. \ref{fig:JapanConfirmed_diffvsChinafirs}) so, it is speculative at best to say that Japan is in its final stages of the outbreak.

\subsubsection*{Spain}
\begin{figure}
\subfloat[ ]{\includegraphics[width =.5\linewidth]{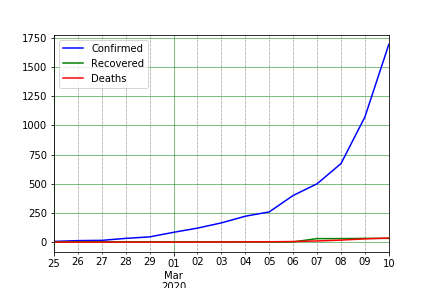}\label{fig:SpainConfirmedvsDeathsvsRecove}} 
\subfloat[ ]{\includegraphics[width = .5\linewidth]{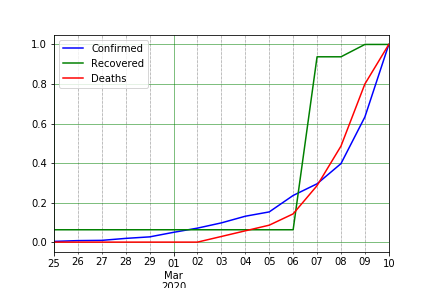}\label{fig:SpainNormalConfirmedvsDeathsvs}}\\
\subfloat[ ]{\includegraphics[width =.5\linewidth]{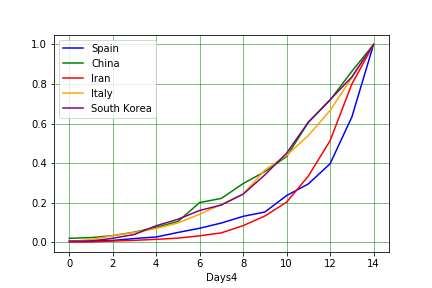}\label{fig:SpainConfirmedvsChinafirst15Ir}} 
\subfloat[ ]{\includegraphics[width = .5\linewidth]{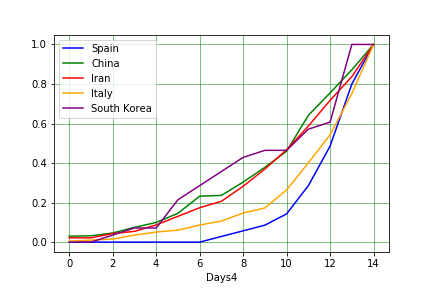}\label{fig:SpainDeathsvsChinafirst15IranI}}\\
\caption{COVID-19 epidemic data and statistics of Spain: a) Confirmed, Recovered and Death COVID-19 cases, b) Normalized data for confirmed, recovered and death COVID-19 cases, c) Confirmed cases of different countries (Spain, China, South Korea, Italy, and Iran), and d) Death cases of the same countries.}
\label{fig:SpainData}
\end{figure}

Massive levels of COVID-19 outbreak were reported around Feb 25 for Spain, as seen by its trend of confirmed, death and recovered cases for demonstrated in Fig. \ref{fig:SpainConfirmedvsDeathsvsRecove}
and its death and recovered cases in Fig.  \ref{fig:SpainNormalConfirmedvsDeathsvs}. By comparing the early stages of Spain with Iran, Italy and South Korea in Fig. \ref{fig:SpainConfirmedvsChinafirst15Ir} the trend since March 10 shows Spain has larger growth rate than other ones mentioned here. Mean age of Spain is 45 years old and hospital beds per person in the country is 29.65 for every 10000 (At 2013), compared to Italy the mean age of Spain is slightly smaller while hospital beds per person are relatively higher, consequently it is expected that the CFR should therefore be slightly lower. After calculating the confirmed to death cases duration which is 3 days the CFR of Spain results in 7\% for Spain, as expected.

\subsubsection*{France}
\begin{figure}
\subfloat[ ]{\includegraphics[width =.5\linewidth]{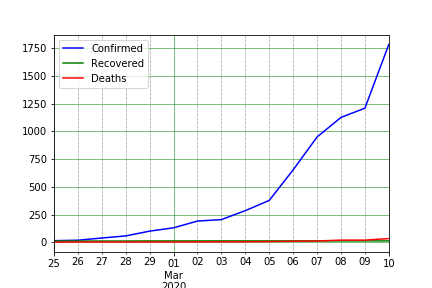}\label{fig:FranceConfirmedvsDeathsvsRecov}} 
\subfloat[ ]{\includegraphics[width = .5\linewidth]{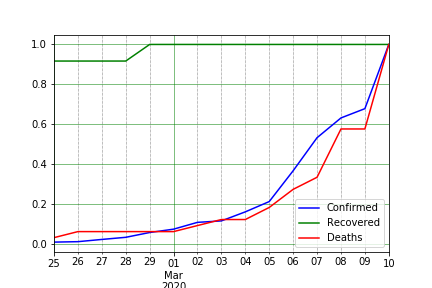}\label{fig:FranceNormalConfirmedvsDeathsv}}\\
\subfloat[ ]{\includegraphics[width =.5\linewidth]{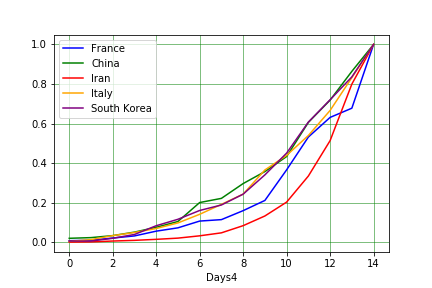}\label{fig:FranceConfirmedvsChinafirst15I}} 
\subfloat[ ]{\includegraphics[width = .5\linewidth]{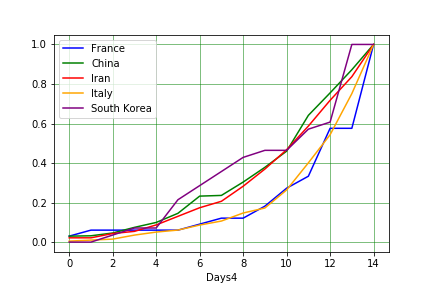}\label{fig:FranceDeathsvsChinafirst15Iran}}\\
\caption{COVID-19 epidemic data and statistics of France: a) Confirmed, Recovered and Death COVID-19 cases, b) Normalized data for confirmed, recovered and death COVID-19 cases, c) Confirmed cases of different countries (France, China, South Korea, Italy, and Iran), and d) Death cases of the same countries.}
\label{fig:FranceData}
\end{figure}

France underwent two increments in its rate of confirmed COVID-19 cases. The first one was on Feb 25, therefore chosen as the starting point of the outbreak. Second was from 3rd to 5th of March as shown in Fig. \ref{fig:FranceConfirmedvsDeathsvsRecov}.
By normalizing data displayed in Fig. \ref{fig:FranceNormalConfirmedvsDeathsv} and calculating the confirmed to death duration which is 3, CFR for France at March 10 results in 3.5\%. Considering the mean age as a parameter, which in France is lower than Spain (42 and 45 respectively) in addition to beds per person which is 64.77 (At 2013) \cite{WHOHBPP}  (Higher than Spain) , CFRs are expected to be lower than Spain. By comparing the growth of confirmed cases with China, Iran, Italy and South Korea depicted in Fig. \ref{fig:FranceConfirmedvsChinafirst15I}
, it's evident that although COVID-19 in France didn't exhibit a fast rate similar to Iran in the first 14 days of the outbreak, it is predicted however that the CFR will increase as much as Italy's.

\subsubsection*{Germany}
\begin{figure}
\subfloat[ ]{\includegraphics[width =.5\linewidth]{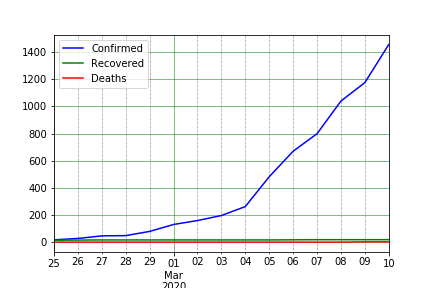}\label{fig:GermanyConfirmedvsDeathsvsReco}} 
\subfloat[ ]{\includegraphics[width = .5\linewidth]{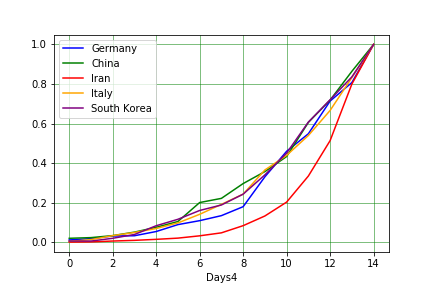}\label{fig:GermanyConfirmedvsChinafirst15}}\\
\caption{COVID-19 epidemic data and statistics of Germany: a) Confirmed, Recovered and Death COVID-19 cases, and b) Confirmed cases of different countries (Germany, China, South Korea, Italy, and Iran)}
\label{fig:GermanyData}
\end{figure}

By observing the Fig. \ref{fig:GermanyConfirmedvsDeathsvsReco} and comparing the trend with China, Italy, Iran and South Korea, which is depicted in Fig. \ref{fig:GermanyConfirmedvsChinafirst15}, Germany had same COVID-19 trend as all others except Iran, which is not a bad news for Germans, but better ones comes when we want to calculate CFR. Total number of deaths in Germany since 10th March is 2. Which is obvious, that the regression approach could not help to find confirmed to death duration. So, like other countries in this stage, we take it as 3 days. By dividing the deaths of 10th march to confirmed of 7th march the 25 out of 10000 (0.25\%) is the CFR, which is too low at this stage.

\subsubsection*{The USA}
\begin{figure}
\subfloat[ ]{\includegraphics[width =.5\linewidth]{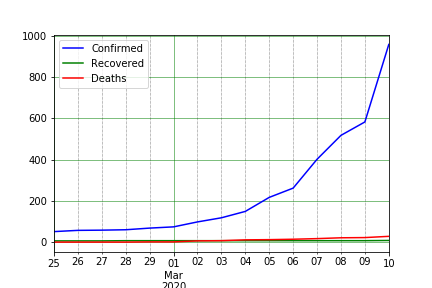}\label{fig:USAConfirmedvsDeathsvsRecovere}} 
\subfloat[ ]{\includegraphics[width = .5\linewidth]{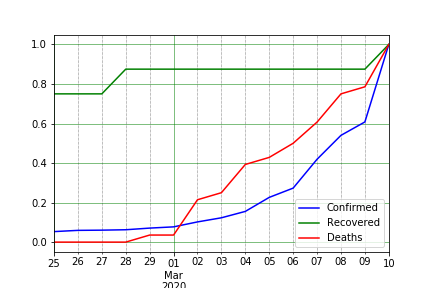}\label{fig:USANormalConfirmedvsDeathsvsRe}}\\
\subfloat[ ]{\includegraphics[width =.5\linewidth]{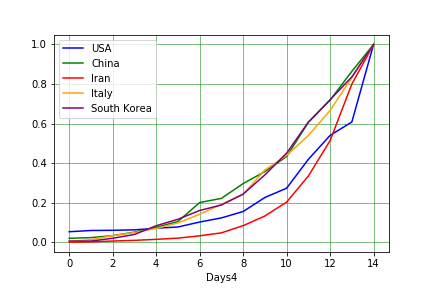}\label{fig:USAConfirmedvsChinafirst15Iran}} 
\subfloat[ ]{\includegraphics[width = .5\linewidth]{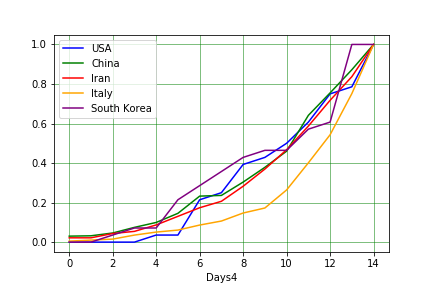}\label{fig:USADeathsvsChinafirst15IranIta}}\\
\caption{COVID-19 epidemic data and statistics of the USA: a) Confirmed, Recovered and Death COVID-19 cases, b) Normalized data for confirmed, recovered and death COVID-19 cases, c) Confirmed cases of different countries (USA, China, South Korea, Italy, and Iran), and d) Death cases of the same countries.}
\label{fig:USAData}
\end{figure}

The USA's confirmed, death and recovered COVID-19 cases trend is depicted in Fig. \ref{fig:USAConfirmedvsDeathsvsRecovere}, as well as its normal form in Fig. \ref{fig:USANormalConfirmedvsDeathsvsRe}.
On March 10 both the confirmed and death cases of the USA increased (about 2.9 and 2.5 times respectively) The recovered cases also increased 2.1 times, but due to its low number of recovered (15 on March 10) we considered it trivial. The huge increase in death and confirmed cases lead the regression line algorithm to define confirmed to death duration in 0 days. However, it wouldn't be possible unless most dead cases are confirmed after death. Accepting the 3 days for the USA, like other countries in this stage, the CFR should be about 14\% which is too high for a country with 38 years old mean age and 29 hospital beds per person. The high CFR could be alluded to the detection of the infected individuals during the last stages of the illness, or reporting on serious cases with higher CFRs exclusively. Compared to China, Iran, Italy and South Korea, both confirmed and death cases of the USA shows higher orders of incrementation. It is advisable to follow the USA in the coming days for more accurate information gathering, considering the latest increment could be caused by wrong data in the previous dates.

%\subsection{Other European}

\section*{Discussion}
\label{sec:Conclusion}
The current outbreak of the novel coronavirus (COVID-19), epi-centered in Hubei Province of the People's Republic of China, has spread to many other countries. The case detection rate is changing daily and can be tracked in almost real time on the mentioned website.\\
Reports on March 10 2020 show that reports show that China has the most confirmed, fatal and also recovered cases and in the terms of confirmed cases, South Korea, Iran, and Italy are following the Chin, respectively. Confirmed death cases lead in numbers by China and followed respectively by Italy, Iran, and South Korea. 
The daily statistics showed that lockdown are effective in reduction of incidence of confirmed cases with COVID-19 after about 11 days in China. South Korea, is one of the first countries reporting the cases after China and the growth pattern of confirmed cases is the same as China's in early stages. However, they implement some policies such as in addition to isolation of people, social avoidance and quarantine policies, and faster detection of infected cases which were effective in decrease in new confirmed case and also case fatality. Italy and China have the approximately same growth rate patterns for first 14 days. Although it is too early to observe quarantines policy's effects in Italy, lockdown strategy of Lombardy (the center of outbreak in Italy) seems to have had a positive effect on other municipalities. Unlike to China's growth pattern, Iran's incremental trend continued to rise until the 20th day of the outbreak, Even Though a temporal reduction in new cases could be observed due to a nation wide
Implemented policy reducing working hours. The case fatality rate in China was 3.8\% (for Hubei was 4.4\%). The highest and the lowest case fatality rate are belongs to Italy (7.9\%)(and South Korea (0.76\%), respectively. Which represent the effectiveness of their policies in control of the COVID-19.\\
Social distancing was one of the most effective policies to control past epidemic disease by limitation human to human transmission and reducing mortality and morbidity(\cite{ahmed2018effectiveness},\cite{caley2008quantifying},\cite{caley2008quantifying}). However studies suggest that combination of multiple policies can boost the effectiveness. For instance, New York City Department of Health implement different policies during the influenza pandemic in 1918-19 at the same time and they have the lowest rate of mortality on the eastern seaboard of the USA\cite{markel2007nonpharmaceutical}.\\
During COVID-19 outbreak, researchers predicted that the mass movement of people at the end of the Lunar New Year holiday, would increase the spreading of disease. Facing this concern, government of China implemented policies which was helpful in control the disease such as, extended the holiday so the holiday would long enough to shelter the incubation period of COVID-19, isolation of confirmed cases in hospitals, quarantining mild or asymptomatic persons in different hospitals, home-based quarantine of people from Hubei province (epicenter of the epidemic), and the most important one was to prevent individuals with asymptomatic infections from spreading the virus.(\cite{li2020early},\cite{chen2020covid})  Iran is facing this concern as an important upcoming event in Iran is Nowruz which is the Iranian New Year, which recommended prompted policies from government.

\bibliographystyle{plain}
\bibliography{Ph.DThesis}
\end{document}